\documentclass[aps,reprint,superscriptaddress,
nofootinbib,
amsmath,amssymb,
prc
]{revtex4-1}
\usepackage{algorithm}
\usepackage{algpseudocode}
\usepackage{natbib}
\usepackage{graphicx}
\usepackage{newtxtext}
\usepackage{dcolumn}
\usepackage{bm}
\usepackage[T1]{fontenc}
\usepackage{ulem}
\usepackage{url}
\graphicspath{{./pic/}}
\usepackage[usenames]{color}
\usepackage[colorlinks=true,allcolors=blue,breaklinks]{hyperref}

\begin{document}

\preprint{...}

\title{Non-parametric Bayesian approach to extrapolation problems \\ in configuration interaction methods}
\author{Sota Yoshida}\email{s.yoshida@nt.phys.s.u-tokyo.ac.jp}
\affiliation{Department of Physics, the University of Tokyo,~Hongo,~Bunkyo-ku,~Tokyo~113-0033,~Japan}
\affiliation{Liberal and General Education Center, Institute for Promotion of Higher Academic Education,~Utsunomiya University,~Mine,~Utsunomiya,~321-8505,~Japan}


\date{\today}

\begin{abstract}
We propose a non-parametric extrapolation method based on constrained Gaussian processes for configuration interaction methods.
Our method has many advantages: (i) applicability to small data sets such as results of {\it ab initio} methods, (ii) flexibility to incorporate constraints, which are guided by physics, into the extrapolation model,
(iii) providing predictions with quantified extrapolation uncertainty, etc.
In the present study, we show an application to the extrapolation needed in full configuration interaction method as an example.
\end{abstract}

\maketitle

\section{Introduction} Recent developments in nuclear potentials based on chiral effective field theory (chiral EFT)~\cite{EMrev1,EGMrev} and {\it ab initio} methods~\cite{QMC_Rev15,NCSMrev16,CCrev,IMSRGrev16,SCGFrev17} have provided deep insights into nuclear many-body problems starting from the fundamental interaction between nucleons, see very recent works~\cite{Lonardoni18,Morris18,Drischler19,Gysbers19} and references therein.

The Full configuration interaction (FCI) method, which is also known as no-core full configuration (NCFC)/no-core shell model (NCSM)~\cite{Barrett_PPNP,Navratil_rev}, is one of the successful {\it ab initio} methods. 
In FCI, wave functions are represented in a truncated sub-space, and the truncation is typically specified by the parameter $N_\mathrm{max}$ which defines the maximum number of harmonic oscillator quanta allowed in the many-body states above the lowest configuration for a target nucleus.

Despite enormous efforts for developing efficient codes~\cite{Caurier_rev,MFDn1,*MFDn2,*MFDn3,KSHELL1,*KSHELL2,BIGSTICK,Forssen18} and advances in computing power,
the currently available $N_\mathrm{max}$ for upper $p$-shell nuclei is around $10$ (see e.g. \cite{Maris_09}) and this is still far from $N_\mathrm{max}=\infty$ corresponding to exact calculations.
One usually extrapolates the sequence of results with different $N_\mathrm{max}$ to $N_\mathrm{max}=\infty$ to estimate the exact value.
In previous studies, several extrapolation methods were proposed and the dependence on them was analyzed~\cite{Zhan04,Maris_09,Coon12,Furnstahl12,More13,Wendt2015,Forssen18}.
The most intuitive example is one based on an exponential function~\cite{Maris_09}.

In addition to FCI, such extrapolation techniques are also required in CI calculations for a valence space,
which is the so-called shell model, using additional truncations.
Representative examples of the truncations are importance truncation scheme~\cite{Roth_IT1,*Roth_IT1b,*Roth_IT1c,Roth_IT2} and Monte Carlo shell model~\cite{MCSMrev1,MCSMrev2}.
In those calculations, the rapid growth of many-body basis for a valence space is alleviated by selecting a small subset of the many-body basis states which is physically more relevant.
These truncation schemes have been successfully applied to valence CI and also FCI calculations in previous works such as~\cite{Roth_IT3,Roth_IT4,Roth_IT5,Abe2012,Liu_NCMCSM,Togashi_MCSM}.

In these studies, extrapolation is performed with some specific functions such as exponential or polynomials, and its coefficients are determined so as to minimize the $\chi^2$ deviation from the given calculated data points.
While any of these offers intuitively reasonable extrapolated results, there is a risk of overfitting, i.e., lack of predictive power for true exact values.
This overfitting is because $\chi^2$ minimization of a parametric model and point estimation of the parameter leads to too deterministic predictions due to the limited expression power of the model.
This is a problematic situation if one intends to discuss quantitative issues like a level ordering of states with small energy differences, the positions of proton and neutron drip lines, and so on.

In the present study, we propose a novel non-parametric extrapolation method for CI calculations using constrained Gaussian processes (GPs), which can give extrapolated results with quantified uncertainty in a systematic manner.

Evaluating extrapolation uncertainties are helpful for breaking down possible origins of discrepancy between FCI results and experimental data, though the major source of uncertainty is, at the moment, from the input potentials: the low-energy constants and the truncated expansion in chiral EFT.
Although GPs are also not free from overfitting, GPs allow us to consider a wider class of functions, and then it is expected to alleviate underestimating extrapolation uncertainties due to the specific choice of the extrapolation function.

We also note that extrapolation techniques using an artificial neural network (ANN) are proposed recently~\cite{IowaANN,OakRidgeANN}.
To train networks, one usually requires large data sets.
However, it is still tough to achieve an enormous number of {\it ab initio}
calculations while varying their inputs such as $N_\mathrm{max}$ and harmonic oscillator parameter $\hbar\Omega$.
In future applications of full CI and also valence CI methods with importance truncation to heavier systems, it is strongly desired to develop an extrapolation technique applicable even to sparse data.
The proposed method is applicable to small data sets too.

We demonstrate the validity of our model by taking the ground state energies obtained by FCI calculations as an example. The code is available on the author's GitHub page~\cite{Supple_SY}.

\section{Formulation of constrained Gaussian Process\label{sec:Formulation}}
Gaussian Process (GP) is a popular statistical method as a non-parametric regression model~\cite{GPML}.
It is also becoming popular in physics due to its flexibility (see e.g. recent applications published in APS journals~\cite{Neufcourt18,Rodrigo18,Siddhartha18,Daheyun19,Neufcourt19,Piekarewicz19,Jordan19}).
The GP regression can be interpreted as a method to describe distribution over a function space and to perform inference of the probability for each function.
This is just what we need because this enables us to consider an ensemble of many possible functions for the extrapolation, infer probability of each sample function, and then quantify uncertainties in the extrapolated value in a statistical manner.

Interestingly, GPs are mathematically equivalent or related to many other models such as ANN, support vector machines, spline models, and so on. We refer the interested reader to e.g.~\cite{Neal95,GPML,DNNGP}.

Here we introduce some notations.
As in statistics literature,  $P(a|b)$ denotes the probability distribution of $a$ under the condition $b$,
and we use $\mathcal{N}(\boldsymbol{\mu},\boldsymbol{\Sigma})$ to express the multivariate Gaussian distribution with mean vector $\boldsymbol{\mu}$ and covariance matrix $\boldsymbol{\Sigma}$.
In what follows, we consider two variable sets, {\it data} and {\it prediction}.
The terminology {\it data}, which is distinguished from {\it experimental data}, is used to express a set of $X = \{x_i|i=1,...,D \}$ and $Y = \{y_i|i=1,...,D\}$. Here we assumed that we have $D$ input points. The {\it prediction} represents positions $X^* = \{x^*_i |i=1,...,P \}$ and values $Y^* = \{y^*_i |i=1,...,P \}$ for $P$ points where the target values are not known.
Especially in our applications, $X$ denotes currently computable $N_\mathrm{max}$, and $X^*$ is a set of $N_\mathrm{max}$ at which FCI calculations have not done (e.g. larger $N_\mathrm{max}$).

In Gaussian processes, it is assumed that the two target values at the two arbitrary points in the vicinity must be {\it similar}, and the so-called kernel functions express the similarities.
Then the data values $Y$ and prediction values $Y^*$ are assumed to be generated from the multivariate Gaussian distribution $\mathcal{N}(\boldsymbol{\mu},\boldsymbol{\Sigma})$ whose covariance matrix $\boldsymbol{\Sigma}$ is given as
\begin{align}
\Sigma = 
\left[
\begin{array}{cc}
K_{XX} & K_{XX^*} \\
K^T_{XX^*}& K_{X^*X^*}
\end{array}
\right].
\label{eq:Kernel}
\end{align}
Here $K_{XX}$, $K_{XX^*}$, and $K_{X^*X^*}$ are respectively $D\times D$, $D\times P$, and $P\times P$ matrices,
and these elements are evaluated with a kernel function.
It is a common practice to choose this kernel function as the radial basis function (RBF)
or the Mat\'ern kernel with $\nu=3/2$ (Mat32) or $\nu=5/2$ (Mat52) (see Appendix~\ref{appA}).

In this work, we use the logMat52 kernel function for the reasons described in Appendix~\ref{appA}.
The logMat52 kernel is defined for, e.g. two data points $x_i$ and $x_j$, as follows:
\begin{align}
k_\mathrm{logMat52}(x_i,x_j) & =  \tau  \left(  
1+\frac{\sqrt{5}r}{\ell} + \frac{5r^2}{3\ell^2}
\right)
\exp{ \left( -\frac{\sqrt{5}r}{\ell} \right)},\label{eq:kernel}
\end{align}
where $r\equiv|x_i-x_j|$, and the global strength $\tau$ and correlation length $\ell$ are the hyperparameters.
Let $\boldsymbol{\theta}$ denote the vector of hyperparameters.
We will revisit the issue of hyperparameters later.

Once the kernel function and its hyperparameters are fixed, one can define the joint covariance matrix $\boldsymbol{\Sigma}$ in Eq.~\eqref{eq:Kernel} for data/prediction as a function of $\boldsymbol{\theta}$.
Then, the joint distribution of data $\boldsymbol{y}$ and predictions $\boldsymbol{y}^*$ under the hyperparameters is given as
\begin{align}
P(\boldsymbol{y},\boldsymbol{y}^*|\boldsymbol{\theta}) &= \mathcal{N}
\left(
\left[
\begin{array}{cc}
\boldsymbol{\mu} \\
\boldsymbol{\mu}^* \\
\end{array}
\right],
\Sigma(\boldsymbol{\theta})
\right).
\label{eq:GPdef}
\end{align}

It is a common practice for mean vectors to be normalized, i.e., $\boldsymbol{\mu}$ and $\boldsymbol{\mu}^*$ are fixed as zero vectors and the data is scaled to have zero mean and unit variance.
The dependence on the choice of mean vectors is also discussed later.

By definition of conditional probabilities, the left-hand side of Eq.~\eqref{eq:GPdef} can be rewritten as
\begin{align}
P(\boldsymbol{y},\boldsymbol{y}^*|\boldsymbol{\theta}) = P(\boldsymbol{y}^*|\boldsymbol{y},\boldsymbol{\theta}) P(\boldsymbol{y}|\boldsymbol{\theta}).
\label{eq:marginal}
\end{align}
Under given $\boldsymbol{\theta}$, one can write down $P(\boldsymbol{y}^*|\boldsymbol{y},\boldsymbol{\theta})$ and $P(\boldsymbol{y}|\boldsymbol{\theta})$ in a closed form:
\begin{align}
P(\boldsymbol{y}^*|\boldsymbol{y},\boldsymbol{\theta}) &
= \mathcal{N}(\boldsymbol{\mu}_{\boldsymbol{y}^*|\boldsymbol{y}} , \Sigma_{\boldsymbol{y}^*|\boldsymbol{y}}), \label{eq:ys_ytheta}\\
\boldsymbol{\mu}_{\boldsymbol{y}^*|\boldsymbol{y}} (\boldsymbol{\theta}) &
= \boldsymbol{\mu}^* + K^T_{XX^*}K^{-1}_{XX}(\boldsymbol{y}-\boldsymbol{\mu}),  \label{eq:mujoint}\\
\Sigma_{\boldsymbol{y}^*|\boldsymbol{y}} (\boldsymbol{\theta}) &=
K_{X^*X^*}-K^T_{XX^*}K^{-1}_{XX}K_{XX^*}, \label{eq:Sjoint}\\
P(\boldsymbol{y}|\boldsymbol{\theta})&=\mathcal{N}(\boldsymbol{\mu},K_{XX}).
\label{eq:likelihood}
\end{align}

Regarding the hyperparameters, the so-called maximum a posteriori (MAP), i.e. one to maximize the hyperparameter posterior $P(\boldsymbol{\theta}|\boldsymbol{y})$, is often used. However, we do not use a single value for the hyperparameters to receive benefit of GPs; various hyperparameters gives us a much wider class of sample functions.
We do inference of their probability distributions by a sampling scheme to integrate out the hyperparameter dependence.
In this case, the posterior distribution of $\boldsymbol{y}^*$ for unobserved input $\boldsymbol{x}^*$ is written as
\begin{align}
P(\boldsymbol{y}^*|\boldsymbol{y}) & \propto  \int P(\boldsymbol{y}^*|\boldsymbol{y},\boldsymbol{\theta})P(\boldsymbol{y}|\boldsymbol{\theta})P(\boldsymbol{\theta}) d\boldsymbol{\theta}.
\label{eq:ypost}
\end{align}
In addition to this, we extend this formulation to more general one to incorporate physics information into GPs.
In many practical situations, the target function is known 
to have shape constraints (e.g. monotonicity or convexity) or inequality constraints.
That is also the case with problems of interest, i.e. energy eigenvalues in FCI are monotonic and (almost) convex with respect to $N_\mathrm{max}$.
In general, the accuracy of a statistical model like GP is improved by including such physics information.
To this end, we extend Eq.~\eqref{eq:ypost} to
\begin{align}
P(\boldsymbol{y}^*|\boldsymbol{y},\alpha,\beta,\ldots) \propto  \int  & P(\boldsymbol{y}^*|\boldsymbol{y},\boldsymbol{\theta})P(\boldsymbol{y}|\boldsymbol{\theta})P(\boldsymbol{\theta}) \times \cdots \nonumber \\
& \times P(\alpha, \beta, \ldots|\boldsymbol{y}^*,\boldsymbol{y}) d\boldsymbol{\theta},
\label{eq:ypost_c}
\end{align}
where $P(\alpha, \beta,\ldots|\boldsymbol{y}^*,\boldsymbol{y})$ is the probability that the constraints $\alpha,\beta,...$ are satisfied under the given $\boldsymbol{y}^*$ and $\boldsymbol{y}$.
The contribution to the integral is determined by the balance among the posterior for the prediction, the likelihood for the hyperparameters, the hyperparameter prior, and the fidelity to the constraints.
This expression is justified when the $\boldsymbol{\theta}$ is independent of the constraints (see Appendix~\ref{appB}).
This is rather general expression, i.e.~constraints can be introduced independently for each problem of interest.

In general, the integration in Eq.~\eqref{eq:ypost_c} cannot be evaluated analytically.
Therefore, some approximation or sampling scheme is required.
We evaluate the integration in Eq.~\eqref{eq:ypost_c} by weighted $N_p$ samples as follows:
\begin{align}
& P(\boldsymbol{y}^*|\boldsymbol{y},\alpha,\beta,\ldots) 
\simeq \sum^{N_p}_{i=1} w^{(i)}P(\boldsymbol{y}^{*(i)}|\boldsymbol{y},\boldsymbol{\theta}^{(i)}), \label{eq:sample_y}\\
& w^{(i)} \equiv  
\frac{
P(\boldsymbol{y},\boldsymbol{\theta}^{(i)}) P(\alpha, \beta, \ldots|\boldsymbol{y}^{*(i)},\boldsymbol{y}) 
}{
\sum^{N_p}_{j=1}  
P(\boldsymbol{y},\boldsymbol{\theta}^{(j)})
P(\boldsymbol{y}^{*(j)}|\boldsymbol{y},\boldsymbol{\theta}^{(j)})
P(\alpha, \beta, \ldots|\boldsymbol{y}^{*(j)},\boldsymbol{y}) 
}.
\label{eq:w}
\end{align}
We employ the particle filtering method~\cite{ParticleFilter}~(also known as Sequential Monte Carlo) as a sampling scheme to evaluate the summation in Eq.~\eqref{eq:sample_y}.
In our particle filtering algorithm, {\it states} $\{\boldsymbol{\theta}^{(i)}$, $\boldsymbol{y}^{*(i)}\}$ are assigned to {\it particles} labeled by $i =1,2, ..., N_p$, and those particles are evolved independently according to the Metropolis-Hastings method
with the so-called resampling scheme; at a certain step of the algorithm, the particles which do not respect the physics constraints are discarded.

\section{Problems of interest\label{sec:problem}}
\subsection{FCI results\label{subsec:FCI}}
\begin{figure}[h]
\centering{
\includegraphics[width=8.6cm]{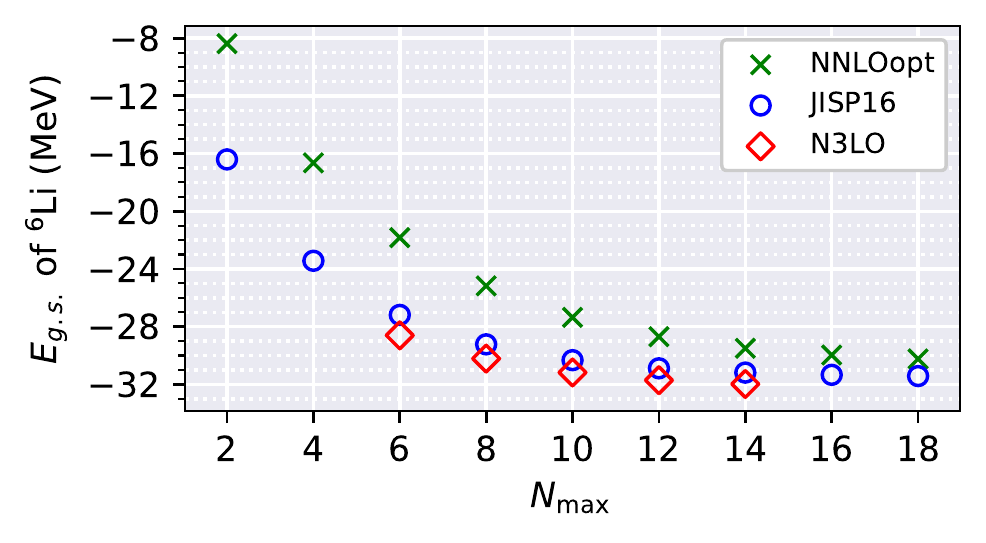}
\caption{The FCI results of g.s. energy of ${}^{6}$Li using JISP16/$\mathrm{NNLO}_\mathrm{opt}$~\cite{Shin_2017} and N3LO~\cite{Kruse_PRC}.
\label{fig:ncsmdata}}
\includegraphics[width=8.6cm]{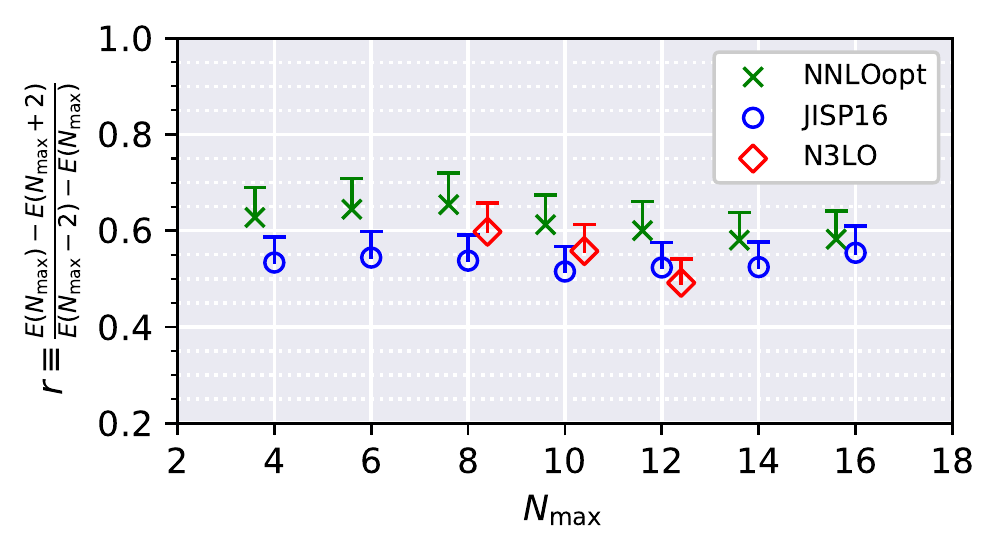}
\caption{The ratios of energy gains associated with 10$\%$ fluctuation (error bar).
The symbols for FCI results are the same as FIG.~\ref{fig:ncsmdata}.
For the visibility of the figure, $\mathrm{NNLO}_\mathrm{opt}$ and N3LO results are slightly shifted to the left and right, respectively.
See the text for more details.
\label{fig:ratio}}}
\end{figure}

In what follows, we apply the constrained GP model to extrapolation problems in FCI calculations.
We analyze published FCI results of the ground state energy of ${}^{6}$Li using JISP16/$\mathrm{NNLO}_\mathrm{opt}$ interaction with $\hbar\Omega=17.5$ MeV~\cite{Shin_2017} and N3LO interaction with $\hbar\Omega=16.0$ MeV which is softened by similarity renormalization group (SRG) method with a flow parameter $\lambda= 2.02$ fm$^{-1}$~\cite{Kruse_PRC}.
The results are summarized in FIG.~\ref{fig:ncsmdata} as a function of $N_\mathrm{max}$.

Let~$\{(x_1,y_1),(x_2,y_2), ..., (x_D,y_D) |x_1<x_2<\cdots<x_D\}$
denote the data, i.e. $(x_1,y_1)=(6,-28.602)$, ..., $(x_D,y_D)=(14,-31.977)$ in the case of N3LO results.
Unlike least squares fitting of parametric models in which one should remove outliers from data,
there is no reason to reduce data in the GP model and we use all $N_\mathrm{max}$ results as data unless otherwise mentioned.

The extrapolation problem addressed below is to estimate the ground state energies at $N_\mathrm{max}$ larger than $x_D$ and we express them as $\{(x^*_1,y^*_1),(x^*_2,y^*_2),...,(x^*_P,y^*_P)|x^*_1 < x^*_2 < \cdots < x^*_P\}$; Here $x^*_1=x_D+2$ and $P$ is large even integer.
For the sake of simplicity, we restrict ourselves to consider the ground state of ${}^{6}$Li with natural parity, i.e. we only consider even $N_\mathrm{max}$.
In practice, we truncate at certain finite $P$ value where predictions are converged with respect to $N_\mathrm{max}$.
A detailed discussion about this $P$ will be given later.

\begin{figure*}
\centering{
\includegraphics[width=17.6cm]{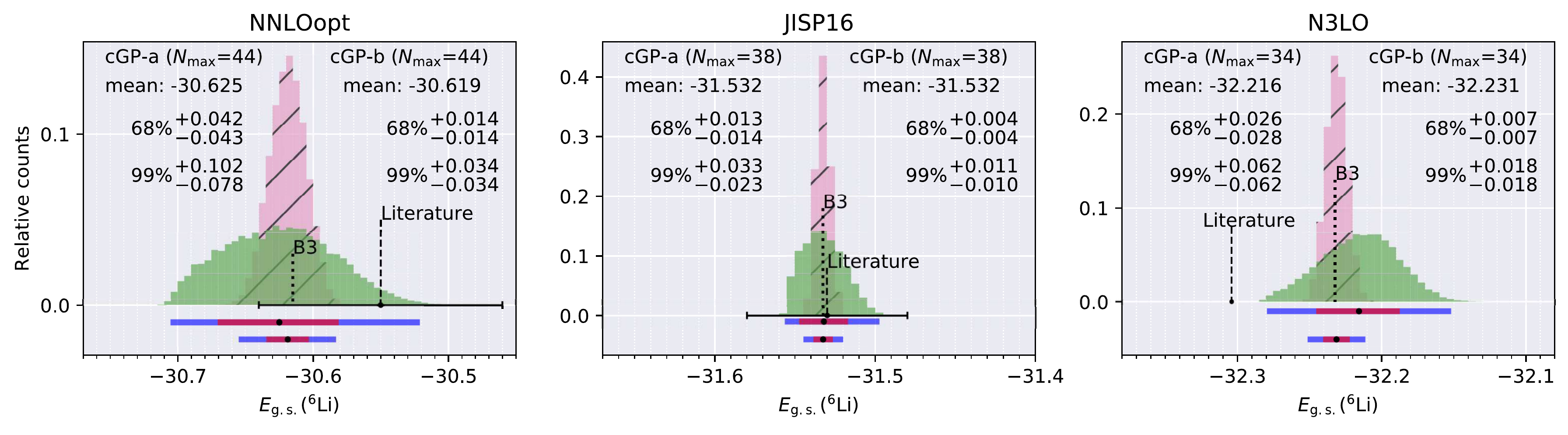}
\caption{Posterior distributions of extrapolated value at certain $N_\mathrm{max}$ are shown by histograms with $5$ keV width.
The B3 (dotted lines) denote the extrapolated values with exponential function and dashed lines are associated with extrapolated values in Ref.~\cite{Shin_2017,Kruse_PRC} with/without error bar.
\label{fig:Extrap_c2}}}
\end{figure*}

\subsection{Constraints on extrapolation function\label{subsec:constraints}}

As minimal constraints on the extrapolation function to capture the asymptotic behavior of FCI results, we impose the following two constraints, $\alpha$ and $\beta$.

The constraint $\alpha$ is variational property, i.e. the monotonicity of energy eigenvalues with respect to $N_\mathrm{max}$:
\begin{align}
P(\alpha|\boldsymbol{y}^*,\boldsymbol{y}) 
 = &\Phi (y_D-y^*_1)\times\Phi (y^*_1-y^*_{2})\times \cdots \nonumber \\
 & \times\Phi (y^*_{P-1}-y^*_P).
\label{eq:monotonicity}
\end{align}
Here we introduced the Probit function
\begin{align}
\Phi(z) &\equiv \int^{z\kappa}_{-\infty}  \frac{1}{\sqrt{2\pi}}\exp{ \left( -\frac{t^2}{2} \right)}  dt,
\end{align}
where $\kappa$ is the controlling parameter of the strictness of constraints.
This probit function approaches the step function at $z=0$ when $\kappa \to \infty$.
This $\kappa$ is set as about $10^6$ in our code, which is large enough to impose the constraints with satisfactory accuracy less than $0.01$ keV.
More precisely, $\kappa$ is gradually increased in our code to $\sim 10^6$ so as to avoid possible localization at the early steps of the Monte Carlo sampling.
We confirmed that the form of this $\kappa$ as a function of the Monte Carlo step does not affect the extrapolation results other than the sampling efficiency.

The second constraint $\beta$ is about the convergence pattern.
We use a ratio of the absolute value of energy gains as a measure of convergence of g.s. energy in FCI.
This ratio $r$ at a certain point $x_j$ is defined as follows:
\begin{align}
r(x_j) \equiv &\left| \frac{y_j-y_{j+1}}{y_{j-1}-y_j} \right|.
\label{eq:ratio}
\end{align}
We plot $\{r(x_j)\}$ for the given data in FIG.~\ref{fig:ratio}.
For the energy eigenvalues by the FCI method, the denominator and numerator in Eq.~\eqref{eq:ratio} are both positive.
If the calculated results of the g.s. energy exactly obey an exponential function, it means $r$ is a constant, which is not the case as shown in FIG.~\ref{fig:ratio}.

This $r$ can be extended to include predictions $\{y^*_j\}$:
\begin{align}
r(x^*_1) &= \left|\frac{y^*_1-y^*_{2}}{y_{D}-y^*_1} \right|, 
r(x^*_2) = \left|\frac{y^*_2-y^*_{3}}{y^*_{1}-y^*_2} \right|, \ldots,\nonumber \\
r(x^*_{P-1}) &= \left|\frac{y^*_{P-1}-y^*_P}{y^*_{P-2}-y^*_{P-1}} \right|, 
\end{align}

We impose the constraint on $\{r\}$ as follows:
\begin{align}
P(\beta|\boldsymbol{y}^*,\boldsymbol{y})
 = &\Phi \left(R_E- r(x^*_1) \right)\times \Phi \left(R_E- r(x^*_2) \right)\times \cdots 
 \nonumber \\
&\cdots \times \Phi \left(R_E- r(x^*_{P-1}) \right),
\label{eq:convergence}
\end{align}
where $R_E$ is an upper threshold determined as follows:
\begin{align}
R_E &= r_\mathrm{mean} + r_\mathrm{std},  \label{eq:RE}\\
r_\mathrm{mean} &\equiv \left| \frac{y_{D-1} - y_D}{y_{D-2}-y_{D-1}}\right|, \label{eq:rmean}\\
r_\mathrm{std} &\equiv \sigma_r r_\mathrm{mean}, \label{eq:rstd}
\end{align}
We use $\sigma_r=0.1$ throughout this work for simplicity.
As can be expected from FIG.~\ref{fig:ratio}, this is a rather moderate constraint on the convergence pattern.
When $\sigma_r$ is large enough, results agree with ones with only the constraint $\alpha$.
We refer to the GP extrapolation model using constraints as the constrained Gaussian process (cGP) model.

\subsection{Choice of the mean function\label{subsec:mean}}

Here we introduce two different mean functions $\boldsymbol{\mu}^{(*)}$ needed in Eqs.\eqref{eq:GPdef}--\eqref{eq:likelihood}:
\begin{itemize}
\item (case a) zero mean: $\boldsymbol{\mu}=\boldsymbol{0}_D, \boldsymbol{\mu}^*=\boldsymbol{0}_P$
\item (case~b) B3 fit: mean of data and prediction are both determined by B3 fit~\cite{Maris_09}, i.e. minimizing $\chi^2$ deviation between the largest three $N_\mathrm{max}$ data and the exponential function in the form of $E_\infty +  c_0\exp{(-c_1N_\mathrm{max})}$ with three free parameters $(E_\infty,c_0,c_1)$. In this choice, it can be said that preliminary knowledge on the behavior of the quantity is included in terms of the mean function of GPs.
\end{itemize}
We refer to these as cGP-a and cGP-b, respectively and analyze both cases below.

\section{Results and Discussions\label{sec:result}}

\begin{figure*}[t]
\centering{
\includegraphics[width=17.2cm]{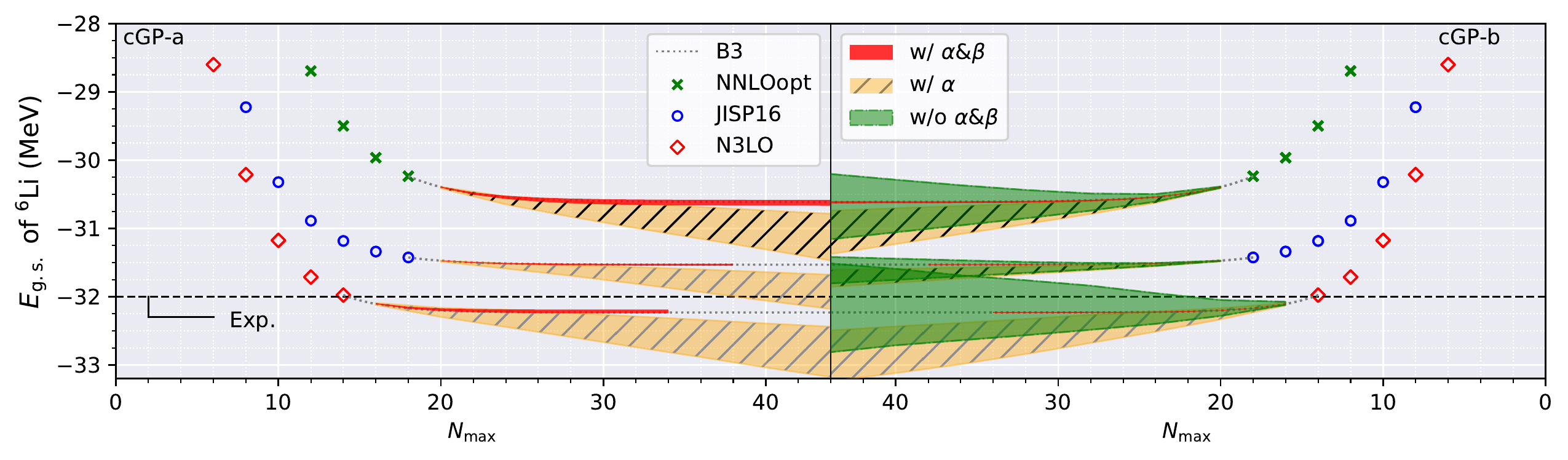}
\caption{The plot showing the impact of the constraints $\alpha$ (monotonicity) and $\beta$ (convexity-like). The bands colored in red, orange (hatched), and green (with dashed dotted line), show the $68\%$ credible intervals of GP predictions with $\alpha$ and $\beta$, only $\alpha$, and without constraints, respectively.
The red bands are cut at a certain $N_\mathrm{max}$ which gives converged results.
}
\label{fig:impact}}
\end{figure*}

\subsection{Extrapolation of g.s. energies \label{subsec:extrap}}

We now present the results of the cGP predictions for the ground-state energies of ${}^{6}$Li in FIG.~\ref{fig:Extrap_c2}.
The extrapolated values at a certain $N_\mathrm{max}$ are shown by histograms.
The cGP-a results are shown by the transparent histograms colored in green, and the hatched histograms colored in pink are for the cGP-b results.

We note that the prediction is truncated up to a certain $N_\mathrm{max}$ where the mean value is converged within 0.2 keV. It means that possible differences between predictions at the finite $N_\mathrm{max}$ and one at $N_\mathrm{max}=\infty$ are suppressed below 1 keV because of the constraint $\beta$.
The $N_\mathrm{max}$ giving converged results are 44,42 and 34 for NNLO$_\mathrm{opt}$, JISP16, and N3LO, respectively.
These numbers are consistent with the intuition that harder interaction requires larger $N_\mathrm{max}$ to obtain converged results, while it should be noted that these results are for different $\hbar\Omega$.
In addition to the mean values, the 68$\%$ and 99$\%$ credible intervals are shown and plotted below the histograms.
In this manuscript, the 68$\%$ interval is determined from the 16th and 84th percentile of the distribution, and the $99\%$ interval is defined in a similar manner.
As a whole, the cGP-b gives smaller credible intervals than the cGP-a.

Other extrapolated values are also shown.
The B3 denotes the exponential fit using the largest three $N_\mathrm{max}$ data.
The conventional B3 extrapolation is always included as a special case of two cGP results.
The values for {\it Literature} are from Ref.~\cite{Shin_2017,Kruse_PRC}.
Here we note that the literature value for N3LO~\cite{Kruse_PRC} might be obtained by an exponential fit using all five data, which is not explicitly stated.
It must also be noted that it is a highly non-trivial task to fairly compare the results with different extrapolation techniques,
because data is truncated in parametric models and some use data with multiple $\hbar\Omega$ as in the A5 extrapolation~\cite{Shin_2017}.

We note that our sampling scheme with the particle filtering gives converged results within a few keV in case of $20,000$ particles after 2,000 times Metropolis-Hastings updates for each particle, and that independent runs reproduce the same results within the Monte Carlo error.

\begin{figure*}[t]
\centering{
\includegraphics[width=8.5cm]{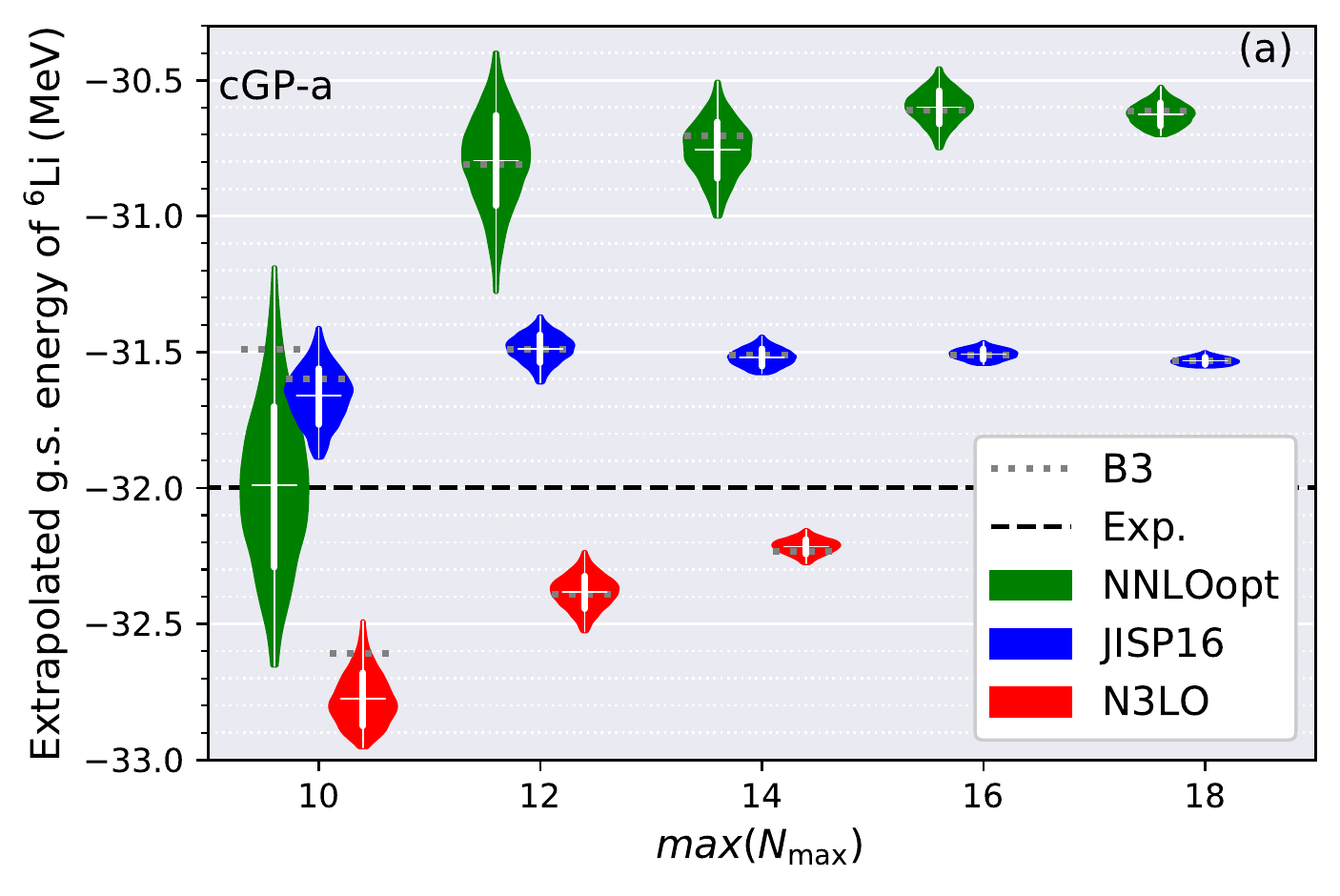}
\includegraphics[width=8.5cm]{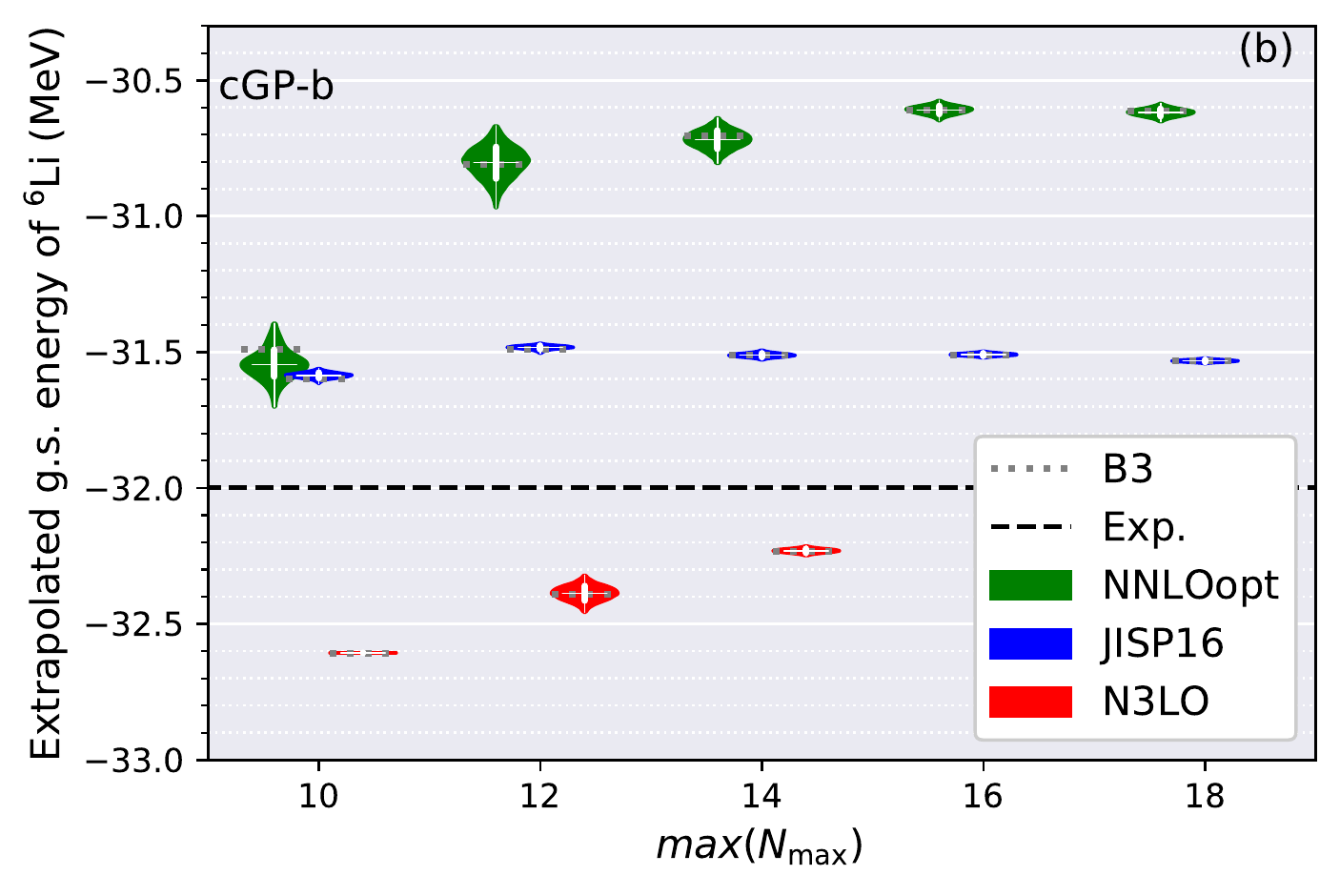}
\caption{Extrapolated ground state energies as a function of maximum $N_\mathrm{max}$ used as data.
The mean values by the cGP models are shown by horizontal line in white.
Shaded areas are obtained by a kernel density estimation of the posterior distribution.
The two types of vertical solid lines, thick and thin ones, show 68$\%$ and 99$\%$ credible intervals, respectively.
For the visibility, the results for $\mathrm{NNLO}_\mathrm{opt}$ and N3LO are slightly shifted to left and right, respectively.
\label{fig:DataDep}}}
\end{figure*}

\subsection{Impact of the constraints\label{subsec:constraints}}

Here we discuss how extrapolation results are influenced by the two constraints imposed.
In FIG.~\ref{fig:impact}, the impact of the constraints is shown.
All symbols are the same as in FIG.~\ref{fig:ncsmdata}, and the cGP-a and cGP-b results are summarized in the left and right regions, respectively.
The bands correspond to the $68\%$ credible intervals of GP predictions with $\alpha$ and $\beta$ (red), with the only $\alpha$ (orange), and without constraints (green), respectively.
Since the $68\%$ errors for the cGP-b prediction is an order of ten keV, the red lines in the right panel are very narrow.

As shown in the textbook~\cite{GPML}, predictions of unconstrained zero-mean GP at points far from the data domain converge to zero with a fixed standard error.
For this reason, we omit the case of cGP-a without $\alpha$ and $\beta$, which is obviously not appropriate for the current purpose.
When one assumes that the wave function is dominated by relatively lower $N_\mathrm{max}$ configurations, predictions with both constraints $\alpha$ and $\beta$ are expected to be more reliable than the others.
 
\subsection{Data dependence \label{subsec:datadep}}

We have used all $N_\mathrm{max}$ results as data so far.
Here we explore the dependence of extrapolated values on the used data to test the potential predictive power of our extrapolation method.
In FIG.~\ref{fig:DataDep}, extrapolated values for both cGP-a and cGP-b are shown as a function of the maximum $N_\mathrm{max}$ used as data, i.e. $x_D=\mathrm{max}(N_\mathrm{max})$.

The mean predictions by the cGP models are shown by horizontal line in white,
and the shaded areas show the posterior distributions obtained by 20,000 particles.
The 68$\%$ and 99$\%$ credible intervals, respectively, are shown by
the thick and thin vertical lines, respectively.
We note that the width of the shaded area is scaled to be the same for each area,
and $\mathrm{NNLO}_\mathrm{opt}$ (N3LO) results are slightly shifted to left (right) for visibility.

As a whole, the size of credible intervals for the cGP-a is larger than that for the cGP-b, and the credible intervals become smaller as higher $N_\mathrm{max}$ data is added with only one exception, which is the cGP-b result for N3LO with max($N_\mathrm{max}=10$).
This exception can be understood from Eqs.~\eqref{eq:marginal}--\eqref{eq:likelihood}.
In this case, the exponential function exactly fits the given three data and then $\boldsymbol{\mu}_{\boldsymbol{y}^*|\boldsymbol{y}}$ in Eq.~\eqref{eq:mujoint} is identical with $\boldsymbol{\mu}^*$.  Any fluctuation of the joint mean value $\boldsymbol{\mu}_{\boldsymbol{y}^*|\boldsymbol{y}}$ is not allowed,
and this significantly reduces probability weights for functions other than the B3 fit.

These plots with quantified uncertainties tell us one criterion of where to stop the massive FCI calculations under the given extrapolation model; It is inadvisable to carry out FCI calculations while increasing $N_\mathrm{max}$ forever, whereas the point to stop must depend on the desired accuracy.

In the rest of this subsection, let us regard the mean values at rightmost max($N_\mathrm{max}$) in FIG.~\ref{fig:DataDep}, i.e. max($N_\mathrm{max}$)=18 for $\mathrm{NNLO}_\mathrm{opt}$ and JISP16, and max($N_\mathrm{max}$)=14 for N3LO, as the (temporary) exact values.

For NNLO$_\mathrm{opt}$ and JISP16, an important remark is that the exact values are covered by the cGP-a predictions with a relatively lower max($N_\mathrm{max}$).
The sign of convergence can be seen around max($N_\mathrm{max}$)=10-12, i.e. one can choose these as points to stop the calculation.
From the behavior of the credible intervals,
the cGP-b seems to underestimate the uncertainties than the cGP-a.
In other words, the cGP-b takes account of fluctuation of the functional form only around the exponential function, while the cGP-a would include a wider class of functions.

For N3LO, any predictions by cGP-a, cGP-b, and B3 fit with lower max($N_\mathrm{max}$) have almost no overlap with the tentative exact value.
It means that all the models fail to estimate the extrapolation uncertainty, while the cGP-a could be better than the others.
The extrapolated values are much more sensitive to the max($N_\mathrm{max}$) than the results with other interactions; the extrapolation for the N3LO results seems to be a more non-trivial problem than the others.
This can be understood from the behavior of the ratio of energy gains $r$. 
Especially in the N3LO case, the $r$ is unstable with respect to $N_\mathrm{max}$, as seen in FIG.~\ref{fig:ratio}.
There are at least two possible explanations for this non-flat behavior of $r$.
One possibility is that the calculation have not yet converged, i.e. the additional bindings by increasing $N_\mathrm{max}$ cannot be regarded as a simple {\it asymptotic} behavior.
The other one is the SRG evolution of the input nuclear potential.
There is not yet enough open data to conclude the origin of the non-flat behavior of $r$.
If one could figure out an additional constraint on the behavior of extrapolated values as a function of max($N_\mathrm{max}$), such a difficulty in extrapolation, which could be observed in particular nuclei and/or particular interactions, would be alleviated.

\section{Summary and Outlook\label{sec:summary}} 

We introduced an extrapolation method for CI-type calculations using constrained Gaussian processes.
This method has the following advantages that are required for future generations of {\it ab initio} studies to make more quantitative discussions on observables of interest and on the quality of adopted nuclear interactions.

Firstly, our extrapolation method does not need to remove {\it outliers} and has applicability to sparse data sets, which are strongly needed for future FCI calculations.
Secondly, one can naturally incorporate domain knowledge into the model.
It is often the case especially in physics that one knows in advance behavior of the target quantity at a certain level, which is ranging from empirical laws to physical principles. One can expect that imposing such information into the extrapolation model improves the accuracy of the predictions.
This flexibility might be useful to alleviate difficulties in the extrapolation for some particular cases.
Thirdly, uncertainty in extrapolation can be quantified in a systematic manner.
Although the main source of uncertainties in FCI calculations is the input nuclear potential,
evaluating extrapolation uncertainties are helpful for further understandings about the nuclear observables.

Regarding uncertainties from input parameters in nuclear many-body methods,
the tremendous efforts to propagate input uncertainty to the observables have been made in the last decade,
see
e.g.~\cite{UQ_Nskin,UQ_RMF,UQ_NskinEDF,
UQ_SkyrmeEDF,Dobaczewski2014,UQ_HFB,
UQ_McDonnell,UQpot_1,*UQpot_1e, UQpot_2,
UQ_CEDF,UQ_chiEFT,UQ_PRX,UQ_opt17,UQ_GSM,SY_UQ}.
In addition to these, it has been shown that eigenvector continuation (EC), which is introduced in Ref.~\cite{EC}, can be used as an efficient emulator of {\it ab initio} methods, then used for uncertainty quantification and sensitivity analyses on input parameters such as low-energy constants in the chiral EFT potentials~\cite{EC2,EC3}.
It is expected that EC (and some other method) facilitates comprehensive studies of uncertainty propagation in {\it ab initio} methods combined with an extrapolation method with quantified uncertainties.

The benefits of the uncertainty quantification are not limited to putting error bars in predictions.
If one properly propagate uncertainties from the input interaction and also quantifies uncertainties such as that due to the extrapolation, it enables us to visualize non-linear relation between input and output of many-body calculations and capability of the many-body method. Then it would provide us with footholds to understand some missing contributions, if there were any.

In this work, we discussed only the ground state energies obtained by FCI calculations.
When it comes to the extrapolation problem of other quantities or in other systems,
the main problem is to find minimal constraints to capture the asymptotic behavior of the quantities well.
It is a possible future direction along this line to extend the cGP model to a higher dimension.
In case of FCI calculations, for example, one can impose the following additional constraint on GP by extending the formulation to $(N_\mathrm{max},\hbar\Omega)$ space: extrapolated values with different $\hbar\Omega$ should converge to the same value to some extent.
The extension of the formulation to a multi-dimensional space is rather straightforward, while it is expected that one needs more technical analyses in numerical studies such as positive semi-definiteness of covariance matrices.
Our model can also be applied to valence CI techniques using an importance-truncation in which extrapolation function is much more non-trivial than the FCI case.
It would also be interesting to apply this kind of cGP to finite-size scaling analyses in other systems.

\newpage

\acknowledgments
The author thanks Noritaka Shimizu for fruitful discussions on future applications to the extrapolation problem in Monte Carlo shell model calculations. This work was supported by JSPS KAKENHI (Grants No. 17J06775).

\normalem

\bibliography{gpextrap_v3}

\appendix

\section{Kernel selection\label{appA}}

\begin{figure}[h]
\centering{
\includegraphics[width=8.5cm]{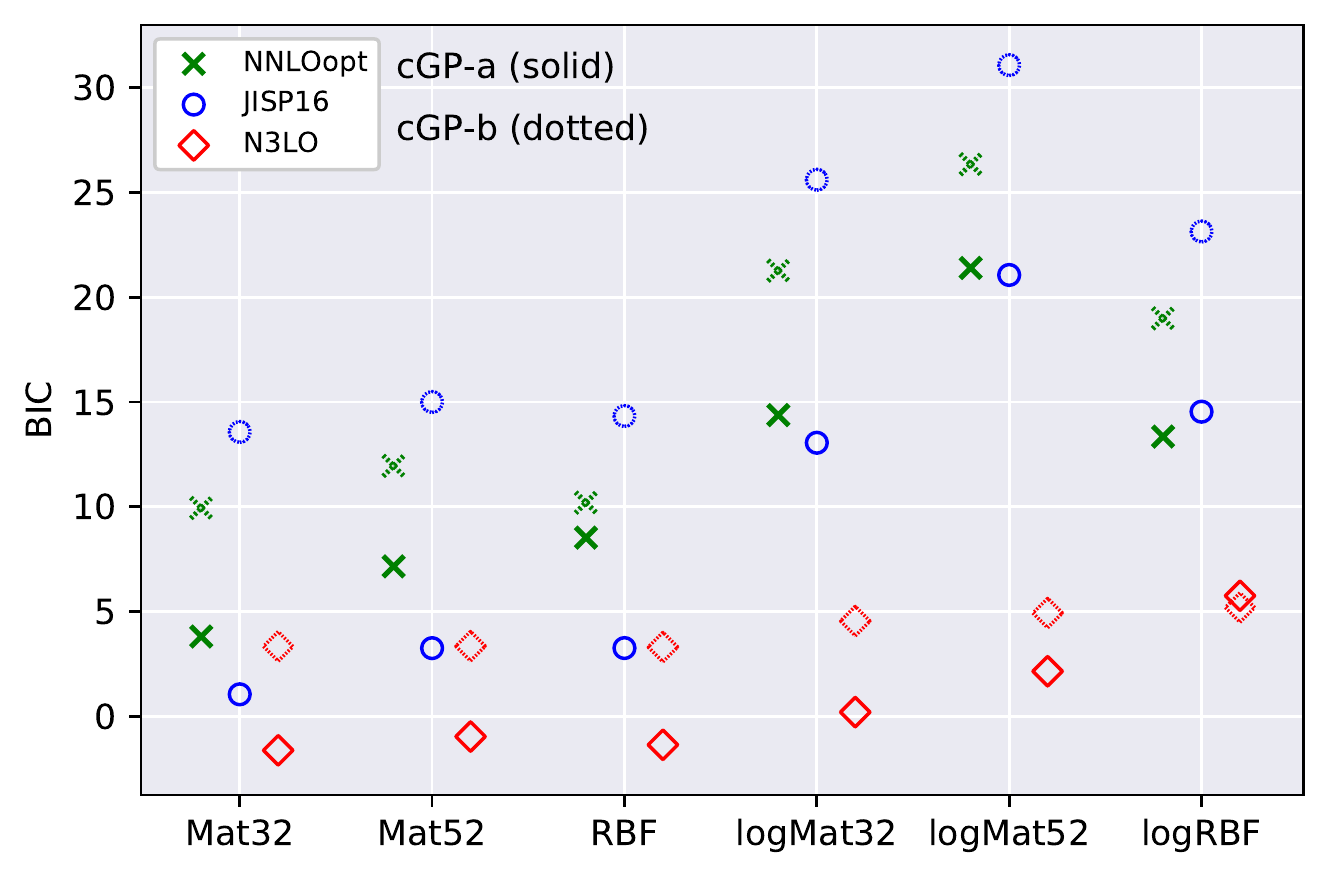}
\caption{The BIC for six kernel functions,
Mat32 (Mat\'ern kernel with $\nu=3/2$),
Mat52 (Mat\'ern kernel with $\nu=5/2$),
RBF (Radial basis function), and their counterparts with logarithm distances.
Results for $\mathrm{NNLO}_\mathrm{opt}$ (N3LO) are slightly shifted to left (right).
Symbols drawn by solid and dotted lines are for cGP-a (zero mean) and cGP-b (exponential mean), respectively.
\label{fig:BIC}}}
\end{figure}

In this section, we discuss the technical details to choose the kernel function.
As mentioned in the main text, popular choices are RBF kernel:
\begin{align}
k_\mathrm{RBF}(x_i,x_j) & = \tau \exp{ \left( - \frac{(x_i-x_j)^2}{2\ell^2} \right)},
\label{eq:RBF}
\end{align}
and Mat\'ern kernel:
\begin{align}
k_\mathrm{Mat\'ern}(x_i,x_j;\nu) & = \tau\frac{2^{1-\nu}}{\Gamma (\nu)} \xi^\nu
K_\nu \left(  \xi \right), \\
\xi &\equiv \frac{\sqrt{2\nu} |x_i-x_j|}{\ell},
\label{eq:Matern}
\end{align}
where $\Gamma$ is the gamma function and $K_\nu$ is the modified Bessel function of the second kind. 
Especially, Mat\'ern kernel with $\nu=3/2$ and $\nu=5/2$ are commonly used.
The RBF kernel corresponds to the special case of the Mat\'ern kernel with $\nu=\infty$.
For the Mat\'ern kernel, a sample function is $k$-times mean square differentiable if and only if $\nu>k$~\cite{GPML}.
For that reason, the Mat\'ern kernels with $\nu > 3/2$ are thought to be appropriate for our purpose, i.e. extrapolation of FCI results, but we include $\nu=3/2$ case too.

\begin{figure*}
\centering{
\includegraphics[width=17.2cm]{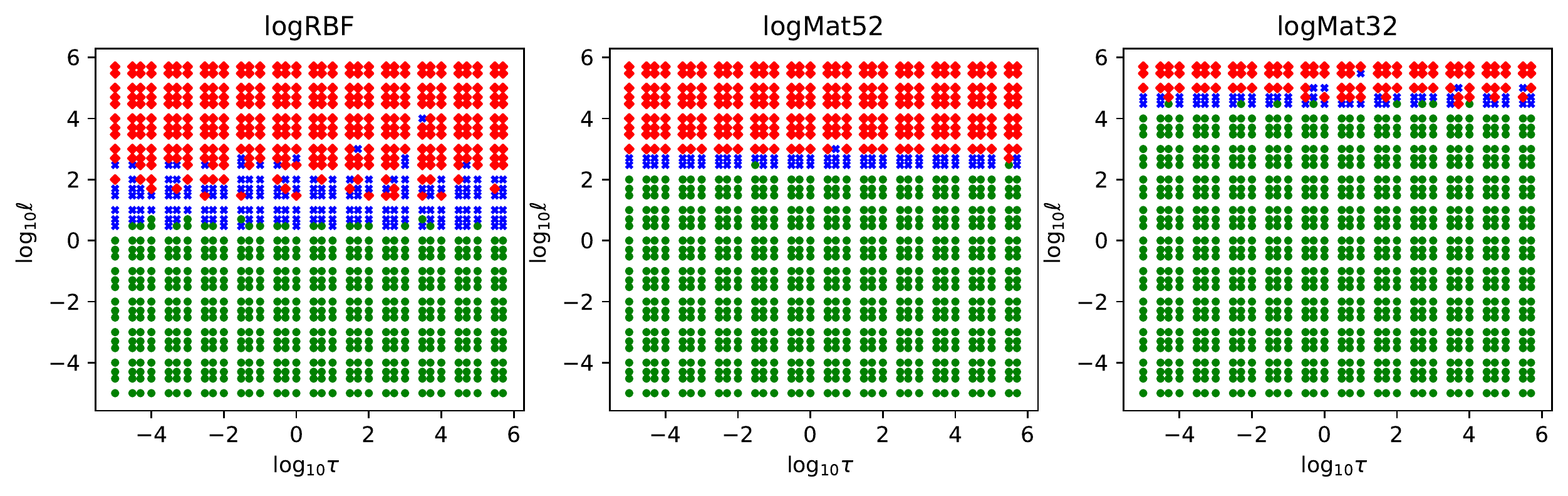}
\caption{
The summary of PSD properties for the three kernel functions.
The covariance matrices having PSD are shown by the filled circle (green).
The diamond symbols (red) mean that both $K_{XX}$ and $\Sigma_{\boldsymbol{y}^*|\boldsymbol{y}}$ are non-PSD, and the cross symbols (blue) are assigned if only the $\Sigma_{\boldsymbol{y}^*|\boldsymbol{y}}$ is non-PSD.
\label{fig:setB}}}
\end{figure*}

\begin{figure*}
\centering{
\includegraphics[width=17cm]{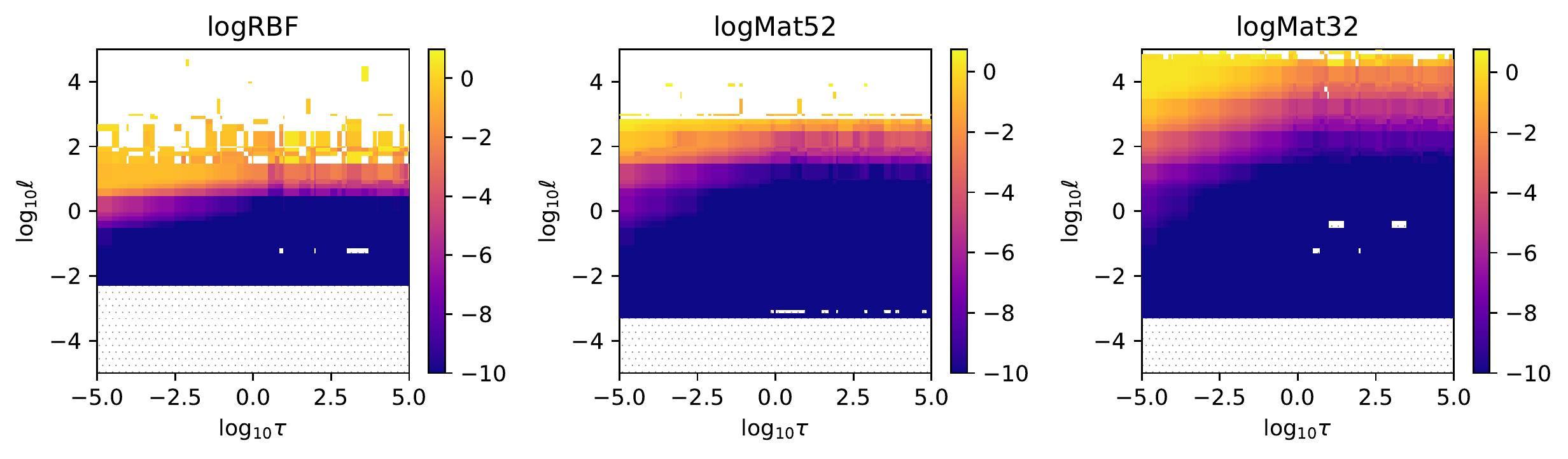}
\caption{The plot showing the impact of the $\epsilon$ prescription to avoid non-PSD.
The colormap shows the $\Delta$ in Eq.~\eqref{eq:Delta}.
The hatched region with dots show the points at which the two codes gives the exactly the same result. The region where our code does not give answer is colored in white.
\label{fig:dif_log}}}
\end{figure*}

In what follows, we consider these Mat32 (Mat\'ern kernel with $\nu=3/2$), Mat52 (Mat\'ern kernel with $\nu=5/2$), RBF (Radial basis function), and their counterparts with logarithm distances:
\begin{align}
k_\mathrm{logRBF} &= \tau \exp{ \left( - \frac{\eta^2}{2\ell^2} \right)},\label{eq:logRBF}\\
k_\mathrm{logMat52} & =  \tau  \left(  
1+\frac{\sqrt{5}\eta}{\ell} + \frac{5\eta^2}{3\ell^2}
\right)
\exp{ \left( -\frac{\sqrt{5}\eta}{\ell} \right)},\label{eq:logMat52}\\
k_\mathrm{logMat32} &=  \tau  \left(  
1+\frac{\sqrt{3}\eta}{\ell} \right)
\exp{ \left( -\frac{\sqrt{3}\eta}{\ell} \right)},\label{eq:logMat32}
\end{align}
where $\eta \equiv |\ln x_i-\ln x_j| $.
Taking the logarithm distance, i.e. replacing $|x_i-x_j|$ in e.g.~Eq.~\eqref{eq:RBF} by $|\ln{x_i}-\ln{x_j}|$, makes results independent on the scale of the x-axis, and allows to capture the non-stationary nature of FCI results (results rapidly converge to certain values as functions of $N_\mathrm{max}$)~\cite{logKernel}.

To compare the six kernels, we here use the Bayesian information criterion (BIC) defined as follows:
\begin{align}
\mathrm{BIC} = \ln P(\boldsymbol{y}|\boldsymbol{\theta}_\mathrm{ML}) - \frac{1}{2} M\ln  n,
\end{align}
where $\boldsymbol{\theta}_\mathrm{ML}$ is the maximum likelihood estimation under the given data, $M$ is the dimension of the hyperparameter $\boldsymbol{\theta}$, and $n$ is the number of data. In this case, the values of the latter term are common among the results with the given interaction.
In FIG.~\ref{fig:BIC}, we summarize the BIC for the six kernels.
Colors and symbols for three interactions are the same as FIG.~\ref{fig:ncsmdata}.
The $\boldsymbol{\theta}_\mathrm{ML}$ is obtained through updates of 2,000 independent particles by the Metropolis-Hastings method,
and the $\boldsymbol{\theta}_\mathrm{ML} $ is converged less than $0.1\%$ accuracy.
As a whole, taking the logarithm of the distance increases the maximum loglikelihood, which is now equivalent to the maximum BIC.

As stated in the main text, we concluded  to use the logMat52 kernel.
This is partly because the logMat52 gives the highest BIC in total.
The other reason is its numerical stability over the (log)RBF kernel. 
The RBF is one of the most popular choices for the kernel.
However, its smoothness of the sample functions is often regarded as too high~\cite{Stein1999},
and, in practice, this too smooth nature sometimes breaks down the positive semi definiteness of covariant matrices in numerical calculations due to rounding errors.
We briefly demonstrate that in the following.

In applications of GPs, one needs to obtain conditional mean vectors Eq.~\eqref{eq:mujoint} and covariance matrices Eq.~\eqref{eq:Sjoint}.
The covariance matrices such as $K_{XX}$ and $\boldsymbol{\Sigma}$ in Eq.~\eqref{eq:Kernel} must be positive semi-definite (PSD) to achieve~e.g.~the Cholesky decomposition for $K^{-1}_{XX}$ and to gemerate samples from the posterior distribution.

In some cases (e.g.~points are located too close to each other), covariance matrices become non-PSD due to rounding errors, although it must be PSD mathematically.
This is also true for our extrapolation method.
The typical prescription to this non-PSD is to add an infinitesimal diagonal matrix to the $K_{XX}$ and/or $\boldsymbol{\Sigma}$.
Let us call this {\it the $\epsilon$ prescription} in the following.
This $\epsilon$ prescription is mathematically equivalent to assume the observation and/or prediction to have noise.

In the current case the target quantity is the calculated energies of light nuclei by {\it ab initio} full configuration interaction method, i.e. in the order of a few tens of MeV.
On the other hand, the typical convergence tolerance of the Lanczos method in shell-model codes on the market is 1.e-5 MeV or better~\cite{KSHELL1,*KSHELL2,BIGSTICK}, i.e. the problems of interest are almost noiseless.

We calculated the conditional mean vectors and covariance matrices while varying the hyperparameters $\tau$ and $\ell$ in Eq.~\eqref{eq:logRBF}--Eq.~\eqref{eq:logMat32}.
In Fig.~\ref{fig:setB}, the positive semi-definiteness of the covariance matrices are summarized.
The diamond symbols (red) correspond to the case that both $K_{XX}$ and $\Sigma_{\boldsymbol{y}^*|\boldsymbol{y}}$ are non-PSD, and the cross symbols (blue) are assigned if only the $\Sigma_{\boldsymbol{y}^*|\boldsymbol{y}}$ is non-PSD.
In terms of the length scale $\ell$, the logRBF gives non-PSD matrices easier than logMat52 by one or two orders of magnitude.

Now we show the effects of the $\epsilon$ prescription on the predictions.
In FIG.~\ref{fig:dif_log}, the following quantity is shown:
\begin{align}
\Delta \equiv \log_{10} \left( max(  
|\boldsymbol{\mu}^\mathrm{own}_{\boldsymbol{y}^*|\boldsymbol{y}} -
\boldsymbol{\mu}^\mathrm{lib.}_{\boldsymbol{y}^*|\boldsymbol{y}}| ) \right),
\label{eq:Delta}
\end{align}
where the superscripts own and lib.~mean the conditional mean vector calculated by our own code and one using the GaussianProcesses.jl library~\cite{GaussianProcesses.jl}, respectively.
For the latter, we fixed the size of observation noise as zero and allowed to use the default $\epsilon$ prescription implemented in the library.
We note that both $\boldsymbol{\mu}^\mathrm{own}$ and $\boldsymbol{\mu}^\mathrm{lib.}$ are now given in a unit of MeV.
The hatched regions with dots for smaller $\ell$ means that the results of the two codes are exactly the same. At the points shown by the diamond symbol in FIG.~\ref{fig:setB}, our code without the $\epsilon$ prescription cannot give the mean vectors. Those correspond to the white region appeared in FIG.~\ref{fig:dif_log}.
When we increase the length scale $\ell$ at which the posterior covariance becomes non-PSD, i.e. points shown by the cross symbol in FIG.~\ref{fig:setB}, the $\Delta$ becomes larger because of the {\it immune system} in the library.
In some cases, the deviations in mean vectors reach a few MeV, which are obviously non-negligible.

One should pay much attention to how the PSD and the observation noise are treated in the codes and whether or not one really can neglect the impact of the $\epsilon$ prescriptions on the predictions, especially when one would like to integrate out the hyperparameter dependence.
For more detailed analyses and codes to reproduce the FIG.~\ref{fig:setB} and FIG.~\ref{fig:dif_log}, we refer the reader to Ref.~\cite{GitHubIssue}.

From the analyses above, we concluded that the logMat52 is the most appropriate choice for our extrapolation method.
\\~\\

\section{Extension of the posterior distribution under the constraints \label{appB}}

Eq.~\eqref{eq:ypost_c} is derived in the followings.
Let the $\boldsymbol{c}$ denote the physics constraints to be imposed.
\begin{align}
P(\boldsymbol{y}^*,\boldsymbol{y},\boldsymbol{c}) & = \int d\boldsymbol{\theta} P(\boldsymbol{y}^*,\boldsymbol{y},\boldsymbol{c},\boldsymbol{\theta}),\\
P(\boldsymbol{y}^*|\boldsymbol{y},\boldsymbol{c}) &= 
\int d\boldsymbol{\theta} \frac{P(\boldsymbol{y}^*,\boldsymbol{y},\boldsymbol{c},\boldsymbol{\theta})}{P(\boldsymbol{y},\boldsymbol{c})},\\
&=
\int d\boldsymbol{\theta} 
\frac{P(\boldsymbol{c}|\boldsymbol{y}^*,\boldsymbol{y},\boldsymbol{\theta})
P(\boldsymbol{y}^*,\boldsymbol{y},\boldsymbol{\theta})
}{
P(\boldsymbol{y},\boldsymbol{c})}
,\\
&\propto
\int d\boldsymbol{\theta} 
P(\boldsymbol{c}|\boldsymbol{y}^*,\boldsymbol{y})
P(\boldsymbol{y}^*|\boldsymbol{y},\boldsymbol{\theta})
P(\boldsymbol{y}|\boldsymbol{\theta})
P(\boldsymbol{\theta})
.
\end{align}

\section{Contributions to the posterior distribution \label{appC}}

\begin{figure*}
\centering{
\includegraphics[width=17.2cm]{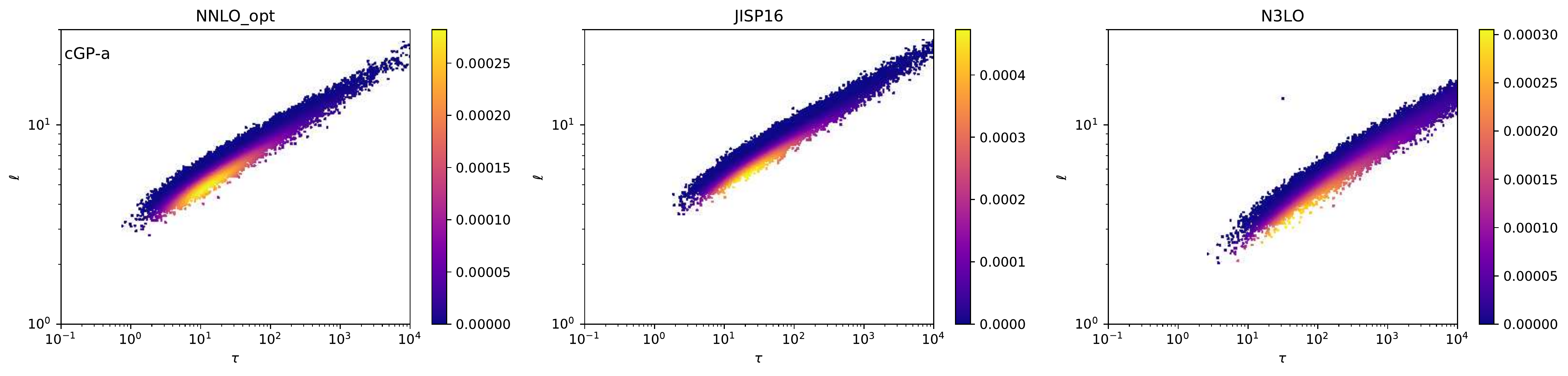}
\caption{Relative contributions to likelihood Eq.~\eqref{eq:likelihood} for cGP-a.
\label{fig:Hyllh-a}}}
\centering{
\includegraphics[width=17.2cm]{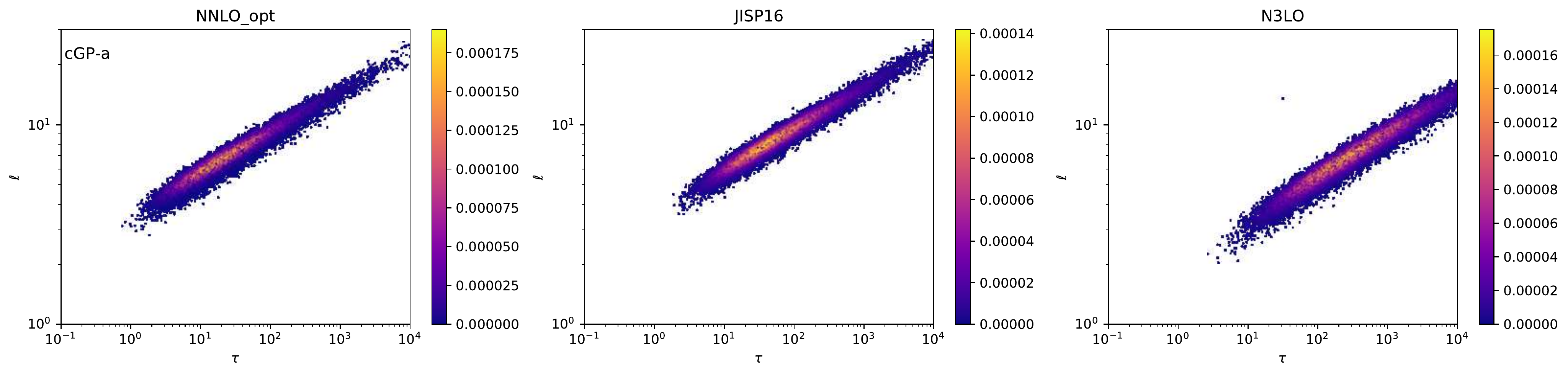}
\caption{Projection plot of the posterior to hyperparameter space.
The colors show relative contributions to Eq.~\eqref{eq:ypost_c} in cGP-a.
\label{fig:Hypos-a}}}
\end{figure*}

\begin{figure*}
\centering{
\includegraphics[width=17.2cm]{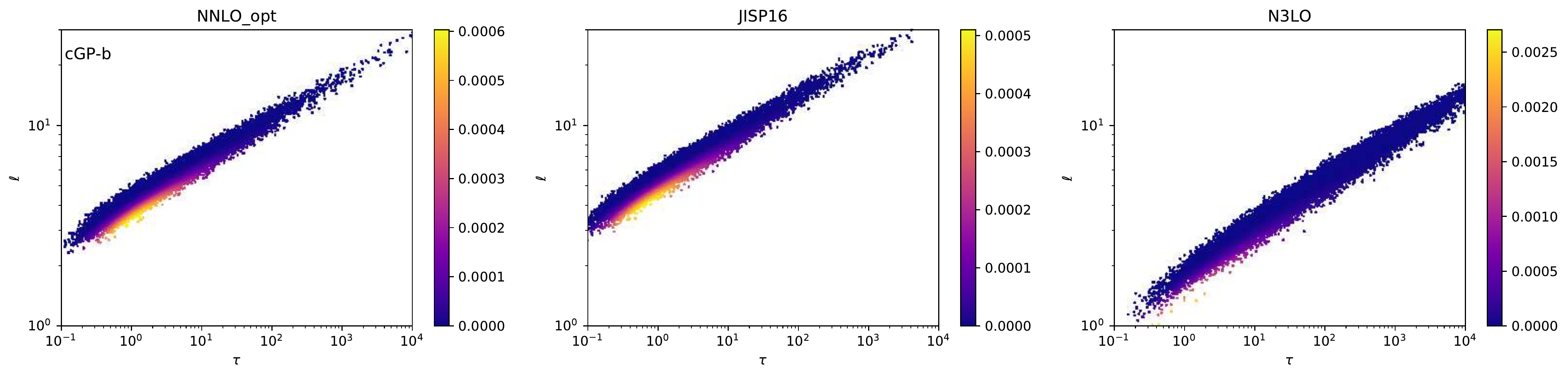}
\caption{Relative contributions to likelihood Eq.~\eqref{eq:likelihood} for cGP-b.
\label{fig:Hyllh-b}}}
\centering{
\includegraphics[width=17.2cm]{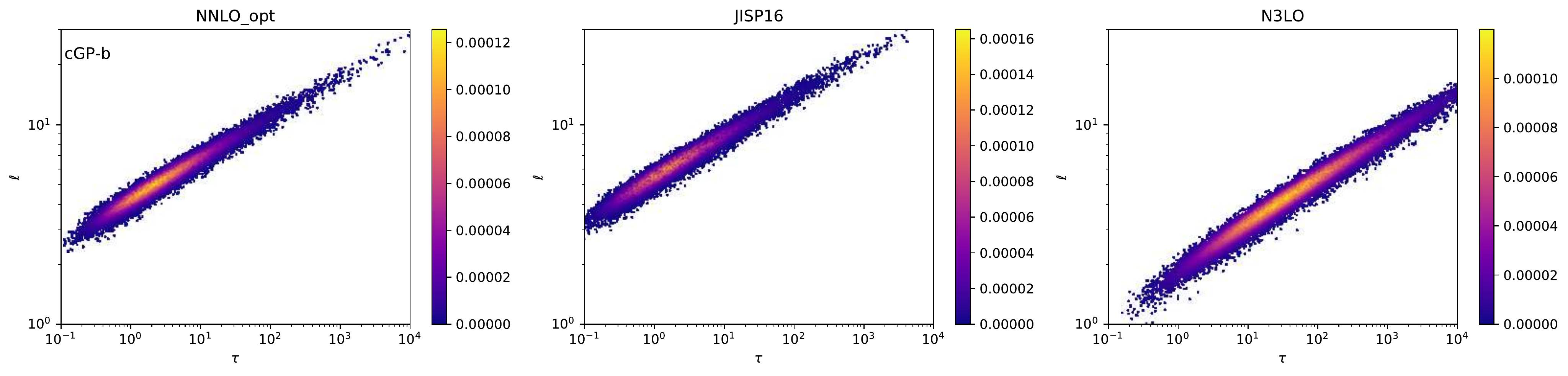}
\caption{Projection plot of the posterior to hyperparameter space.
The colors show relative contributions to Eq.~\eqref{eq:ypost_c} in cGP-b.
\label{fig:Hypos-b}}}
\end{figure*}

In Figures~\ref{fig:Hyllh-a}--\ref{fig:Hypos-b}, the hyperparameter distributions are shown.
Colors of each dot point show the relative size of contributions to the likelihood $P(\boldsymbol{\theta}|\boldsymbol{y})$ and to the integral in Eq.~\eqref{eq:ypost_c}.
In our sampling method, particles are evolved according to the random walk Metropolis-Hastings method.
We can see from FIG.~\ref{fig:Hypos-a} and FIG.~\ref{fig:Hypos-b} that particles distribute like an ellipse in hyperparameter space.
The particles do not distribute over the non-PSD region discussed in Sec.~\ref{appB}.

\newpage

\end{document}